\documentclass[%
 reprint,
  superscriptaddress,
showpacs,
showkeys,
preprintnumbers,
nofootinbib,
 amsmath,amssymb,
aps,
 prd,
]{revtex4}

\usepackage{graphicx}
\usepackage{dcolumn}
\usepackage{bm}
\usepackage{hyperref}

\usepackage{subfigure}
\usepackage{multirow}

\usepackage{epsf}

\def\beq{\begin{equation}}
\def\be{\begin{equation}}
\def\eeq{\end{equation}}
\def\ee{\end{equation}}
\def\bea{\begin{eqnarray}}
\def\eea{\end{eqnarray}}

\newcommand{\eqn}[1]{eq.~(\ref{#1})}
\newcommand{\Eqn}[1]{Eq.~(\ref{#1})}
\newcommand{\eqns}[2]{eqs.~(\ref{#1})-(\ref{#2})}

\newcommand{\gsim}{\mbox{${~\raise.25em\hbox{$>$}\kern-.70em
\lower.25em\hbox{$\sim$}~}$}}
\newcommand{\lsim}{\mbox{${~\raise.25em\hbox{$<$}\kern-.70em
\lower.25em\hbox{$\sim$}~}$}}

\begin{document}


\title{Squeezing out predictions with leptogenesis from SO(10)}

\preprint{DSF-3-2012}

 \bigskip

 \author
{Franco Buccella}
\email{buccella@na.infn.it}
\affiliation
{INFN, Sezione di Napoli, \\
Complesso Universitario di Monte Sant'Angelo,  Via Cintia, 80126 Napoli,
Italy}
 \author
{Domenico Falcone}
\email{domenicofalcone3@virgilio.it}
\noaffiliation
 \author
{Chee Sheng Fong}
\email{chee.sheng.fong@lnf.infn.it}
 \author
{Enrico Nardi}
\email{enrico.nardi@lnf.infn.it}
\affiliation
{INFN, Laboratori Nazionali di Frascati, \\
                Via Enrico Fermi 40,    I-00044 Frascati, Italy}
 \author
{Giulia Ricciardi}
\email{giulia.ricciardi@na.infn.it}
\affiliation
{Dipartimento di Scienze Fisiche, Universit\`a di Napoli Federico II\\
Complesso Universitario di Monte Sant'Angelo, Via Cintia, 80126 Napoli,
Italy}
\affiliation
{INFN, Sezione di Napoli, \\
Complesso Universitario di Monte Sant'Angelo,  Via Cintia, 80126 Napoli,
Italy}

\begin{abstract}
  We consider the see-saw mechanism within a non-supersymmetric
  SO(10) model.  By assuming the SO(10) quark-lepton symmetry, and
  after imposing suitable conditions that ensure that the right-handed
  (RH) neutrino masses are at most mildly hierarchical (compact RH
  spectrum) we obtain a surprisingly predictive scenario. The absolute
  neutrino mass scale, the Dirac and the two Majorana phases of the
  neutrino mixing matrix remain determined in terms of the set of
  already measured low energy observables, modulo a discrete ambiguity
  in the signs of two neutrino mixing angles and of the Dirac
  phase. The RH neutrinos mass spectrum is also predicted, as well as
  the size {\it and sign} of the leptogenesis CP asymmetries. We
  compute the cosmological baryon asymmetry generated through
  leptogenesis and obtain the correct sign, and a size compatible with
  observations.
\end{abstract}

\keywords{Grand Unified Models, Leptogenesis}

\pacs{
12.10.Dm, 
11.30.Fs, 
98.80.Cq, 
13.35.Hb  
}

   \maketitle
\flushbottom

\section{Introduction}

In the last decade, experiments with solar, atmospheric, reactor and
accelerator neutrinos have provided compelling evidences for neutrino
oscillations~\cite{pontecorvo}, which imply nonvanishing neutrino
masses and mixings.  Neutrino oscillation experiments have been
measuring with increasing precision the values of the mixing angles
and of the mass-squared differences.  The value of the absolute neutrino
mass scale is still unknown; however, existing limits imply that this
scale is bafflingly small, much smaller than those of all the other
elementary fermions. The most popular explanation for the neutrino
mass suppression is undoubtedly provided by the see-saw
mechanism~\cite{grs} which requires the existence of very heavy
right-handed (RH) Majorana neutrinos. Fermions with quantum numbers of
RH neutrinos, that are singlets under the standard model (SM) gauge
group, are found in the spinorial {\bf 16} representation of
SO(10)~\cite{georgi,Fritzsch:1974nn}, which therefore provides a
quite natural Grand Unified Theory (GUT) framework to embed the see-saw.

The see-saw RH neutrinos also play a key role in
leptogenesis~\cite{fuya,leptoreview}, which is a very appealing
scenario to explain the origin of the Baryon Asymmetry of the Universe
(BAU).  In leptogenesis, the cosmological baryon asymmetry is seeded
by an initial asymmetry in lepton number generated in the
out-of-equilibrium decay of the RH neutrinos, that is then transferred
in part to baryons by means of the $B+L$ violating `sphaleron'
interactions, that are non-perturbative SM processes.  In SO(10),
the order of magnitude of the RH neutrino masses is fixed around the
scale of the spontaneous breaking of the $B-L$ U(1) symmetry,  and it
is consistent with the values of the RH neutrino masses required for
successful leptogenesis $M_R \sim 10^{11\pm 2}\,$GeV.  Indeed, the
double role of RH neutrinos in the see-saw and in leptogenesis
underlines the importance of deriving information on their mass
spectrum.

Recently, an analysis of the relations between the left-handed (LH)
neutrino observables (mass-squared differences and mixings) and the RH
neutrino spectrum, constrained to be of a compact form (i.e. with
masses all of the same order of magnitude) was carried out within the
framework of an SO(10)-inspired model~\cite{Buccella:2010jc}, and a
scenario for baryogenesis via leptogenesis was also constructed.  The
study in Ref.~\cite{Buccella:2010jc} was carried out under the
simplifying assumption of a vanishing value of the lepton mixing angle
$\theta_{13}$, and in the leptogenesis analysis all lepton flavour
effects~\cite{ flavour0, flavour1, flavour2} as well as the effects
from the heavier RH neutrinos~\cite{N2,Antusch:2010ms} had been
neglected.  However, recent experimental results hint to a nonvanishing value of
$\theta_{13}$~\cite{Schwetz:2011zk,Fogli:2011qn,GonzalezGarcia:2010er}
and imply that the assumption $\theta_{13}=0$ should be dropped. The
inclusion of flavour effects is also mandatory when leptogenesis
occurs below $T\sim 10^{12}\,$GeV, since the one flavour
`approximation' is known to give unreliable results.  Moreover, in the
case of a compact RH spectrum, that is when all the RH neutrino masses
fall within a factor of a few, to obtain a trustworthy result it is
also necessary to include the asymmetry production and washouts from
the two heavier RH neutrinos.

In the present paper we consider a scenario similar to the one in
Ref.~\cite{Buccella:2010jc} improving on several points.  We fix
$\theta_{13}$ to the nonvanishing best fit value given in
Ref.~\cite{GonzalezGarcia:2010er}. This in turn implies that the Dirac
phase $\delta$ of the Pontecorvo, Maki, Nagakawa and Sakata (PMNS)
mixing matrix~\cite{Maki:1962mu,Bilenky:1978nj} enters all the
equations, and in particular contributes to the leptogenesis CP
asymmetries.  Most importantly, we clarify how the conditions ensuring
a compact RH neutrino spectrum have consistent solutions only for
$\delta\neq0$, and how the corresponding solutions yield a
surprisingly predictive scenario in which all the yet unknown low
energy parameters, namely the LH neutrino mass scale $m_{1}$ and the
three PMNS CP violating phases $\delta,\,\alpha$ and $\beta$, remain
determined in terms of already measured quantities, modulo a few signs
ambiguities.  In the high energy sector, the RH neutrino spectrum is
also predicted.  The crucial test of the scenario is then the
computation of the baryon asymmetry yield of leptogenesis.  We include
lepton flavour effects \cite{ flavour0, flavour1, flavour2} in our
analysis and argue that they are crucial to evaluate correctly the
baryon asymmetry. Most importantly, the high level of predictability of
our framework allows to predict both the size and the {\it sign} of
the BAU. By requiring agreement with observations, we are then able to
solve almost completely the residual signs ambiguities.

The paper is organized as follows.  In Section~\ref{fram} we describe
the SO(10) framework and spell out the quark-lepton symmetry
assumption.  In Section~\ref{motiv} we discuss the constraining
conditions that ensure a compact spectrum for the RH neutrinos. In
spite of quark-lepton symmetry a compact form is achieved, and with a
sufficiently large scale not to conflict with the Davidson-Ibarra
bound that, within the $SO(10)$ see-saw, often vetoes successful
baryogenesis via leptogenisis.
In Section~\ref{low} we confront our scenario with
the set of measured low energy neutrino observables, and we work out
predictions for the absolute scale of neutrino masses $m_{1}$, for the
PMNS CP violating phases $\delta,\,\alpha,\,\beta$ and for the RH
neutrino mass matrix.  In Section~\ref{LEPTO} we calculate the various
CP asymmetries in RH neutrino decays, we briefly discuss the procedure
followed to estimate the baryon asymmetry yield of leptogenesis and
stress how a proper treatment of flavour effects is crucial for
obtaining  reliable estimates for the different cases.  Finally
in Section~\ref{CONCLU} we discuss our results and draw the conclusions.

\section{$\mathbf{SO(10)}$  GUT and quark-lepton symmetry}
\label{fram}

We work in a non-supersymmetric grand unified SO(10) model.  We assume
three fermion families whose left-handed (LH) states are assigned to a
${\bf 16}$ spinorial representation of SO(10), which thus contains all
the SM fermion and antifermion states of the same chirality.  In
addition, the ${\bf 16}$ includes one $SU(2)$ singlet neutrino for
each family.  All elementary fermions of opposite chirality (RH) are
assigned to the conjugate representation $\overline{\mathbf{16}}$.
Fermion masses are generated by Yukawa terms of the form:
\beq
Y_{ij} \cdot \mathbf{16}_i\, H^\dagger\, \mathbf{16}_j + {\rm h.c.}
\label{Yuk}
\eeq
where $Y$ is a $3\times 3$ matrix of Yukawa couplings with indexes in
family space, and $H$ denotes a multiplet of scalar (Higgs) bosons.
The tensor product of the fermion representations in eq.~(\ref{Yuk})
gives \beq {\bf 16}\times{\bf 16}={\bf 10}_s +{\bf 126}_s + {\bf 120}_a \eeq
where
the subscripts $s,a$ refer to the symmetric and antisymmetric nature
of the representation in the family indexes.  Thus, to make the Yukawa term in
eq.~(\ref{Yuk}) an SO(10) singlet, $H$ must be assigned to a
$\bf{10}$, to a $\bf{126}$, or to a $\bf{120}$. Clearly, for $\bf{10}$
and $\bf{126}$, that match the two symmetric fragments of the tensor
product, the Yukawa matrix $Y$ is symmetric, while for the $\bf{120}$
it is antisymmetric.

In the present work we consider Yukawa terms involving only the
$\bf{10}$ and $\bf{126}$ that are already needed for the gauge
symmetry breaking pattern (the $\bf{126}$ is also needed to generate
Majorana masses for the see-saw RH neutrinos) and we exclude Yukawa
couplings with the $\bf{120}$ which would imply a departure from
minimality.

We adopt the following  SO(10) breaking  pattern; we also indicate
the set of Higgs $SO(10)$ representations needed for each step
\begin{eqnarray}
SO(10) &\xrightarrow{{\bf 210}} &  SU(2)_L \otimes SU(2)_R \otimes SU(4)\nonumber \\
&\xrightarrow{{\bf 126}}& SU(2)_L \times U(1)_Y \times SU(3)_C  \nonumber \\
&\xrightarrow{{\mathbf{126},\>\, \mathbf{10} }  }&  SU(3)_C \times U(1)_Q
\end{eqnarray}
The first step in this chain is a breaking to a maximal subgroup of $SO(10) $,
the intermediate Pati-Salam group  $SU(2)_L \otimes SU(2)_R
\otimes SU(4)$\cite{patisalam}.
Let us list explicitly useful branching rules
for $SO(10)\to SU(2)_L \otimes SU(2)_R \otimes SU(4)$:
\begin{eqnarray}
  \label{eq:branchings} \nonumber
  \mathbf{210} &\supset&(\mathbf{1},\mathbf{1},\mathbf{1})\oplus \dots
\\  \nonumber
  \mathbf{126} &=&
(\mathbf{1},\mathbf{1},\mathbf{6}) \oplus
(\mathbf{3},\mathbf{1},\mathbf{10}) \oplus
(\mathbf{1},\mathbf{3},\mathbf{\overline{10}}) \oplus
(\mathbf{2},\mathbf{2},\mathbf{15}) \\  \nonumber
  \mathbf{16} &=&   (\mathbf{2},\mathbf{1},\mathbf{4}) \oplus
(\mathbf{1},\mathbf{2},\mathbf{\overline{4}}) \\
  \mathbf{16} \otimes   \mathbf{16}
&\supset&   (\mathbf{1},\mathbf{3},\mathbf{\overline{10}}) \oplus \dots
\end{eqnarray}
The $ SU(2)_L \otimes SU(2)_R \otimes SU(4)$ singlet in the
$\mathbf{210}$ is responsible for the first breaking at the GUT scale.

The $ \mathbf{126}$ then breaks $SU(2)_L \otimes SU(2)_R \otimes SU(4)
\to SU(2)_L\otimes U(1)_Y\otimes SU(3)_C$ at an intermediate scale
$\Lambda_R$, that we assume to be around $ 10^{11}\,$GeV.  The
relevant component that triggers the breaking is
$(\mathbf{1},\mathbf{3},\mathbf{\overline{10}})$ since the
$\mathbf{10}$ of $SU(4)$ contains an $SU(3)$ singlet. The other
components, $(\mathbf{3},\mathbf{1},\mathbf{10}) $ and
$(\mathbf{2},\mathbf{2},\mathbf{15})$ would break $SU(2)_L$, while
$(\mathbf{1},\mathbf{1},\mathbf{6})$ would break colour. The RH neutrino
$N$, together with all the other SM $SU(2)_L$ singlet fields, is
contained in $(\mathbf{1},\mathbf{2},\mathbf{\overline{4}})$, and thus
the bilinear $N\cdot N$ belongs to the fragment displayed in the last
line, which is the only one suited to build up a gauge invariant term
when coupled to the intermediate gauge symmetry breaking component
$(\mathbf{1},\mathbf{3},\mathbf{10})$ of the $\mathbf{\overline{126}}$
Higgs multiplet.  We can also decompose $SO(10) $ according to $SO(10)
\supset SU(5)\,\otimes\,U(1) \supset SU(5)$. With respect to $SU(5)$:
\beq
\mathbf{126} = \mathbf{1}+ \mathbf{\overline{5}}+ \mathbf{10}+
\mathbf{\overline{15}}+ \mathbf{45}+ \mathbf{\overline{50}}\,.
\eeq
Of these representations only the $\mathbf{1}$,
$\mathbf{\overline{5}}$, and $\mathbf{45}$
have neutral colour singlet Higgs components that can have nonzero
vacuum expectation values if $U(1)_{em}\times SU(3)_C$ has to remain
unbroken.  $N$ is a singlet with respect to $SU(5)$ and so it is
$N\cdot N$; therefore, it is the $SU(5)$ singlet in
$\mathbf{\overline{126}}$ that couples to $N\cdot N$ and gives an
invariant mass of the order of the intermediate scale to $N$.  In
terms of the intermediate representations, it is
$(\mathbf{1},\mathbf{3},\mathbf{\overline{10}})$ in last line of
Eq. (\ref{eq:branchings}), that must contain an SU(5) singlet.

As regards the fermion masses, if they originate only from vacuum
expectation values (vevs) of scalars in the $\bf{10}$, the following
relations hold:
\beq
m_D= m_u,    \qquad \qquad m_e= m_d,
\label{prima}
\eeq
where $m_D$ is the neutrino Dirac mass matrix, and $m_u$,
$m_d$ and $m_e$ are respectively the mass matrices for the up and down
quarks and charged leptons.  The two relations in \eqn{prima} are
sometimes referred to as quark-lepton symmetry; they imply for each
generation the GUT scale prediction $m_{e_i}/m_{d_i}=1 $ (with
$i=1,2,3$ a generation index) which however, after including
renormalization group corrections, agrees with observations only for
the third generation ($b$-$\tau$ unification) but is badly violated
for the first and second generations. If instead quark and lepton
masses originate from one or more $\bf{126}$ (that however should be
different from the $\bf{126}$ that breaks the gauge symmetry) the
following relations hold:
\beq
\label{seconda}
m_D= -3 m_u,  \qquad \qquad m_e= -3 m_d\,.
\eeq
The factor of $-3$ is a colour factor between leptons and quarks;  it is reminiscent of the Georgi and Jarlskog
mechanism~\cite{Georgi:1979df}, which in SU(5), when it appears in a
family dependent way,  allows to circumvent the prediction of
unification for the first two families  yielding $m_\mu/m_s\neq m_e/m_d \neq 1$
 while preserving  $m_\tau/m_b=1$.
In SU(5) the discrepancy with the observed values of the down-quarks
and charged lepton masses can in fact be weakened by assuming that the
Yukawa coupling of the second generation to itself involves a $
\mathbf{45}$ of SU(5) scalars, instead of the usual $\mathbf{\bar
  5}$,
yielding mass ratios \beq |m_\mu/m_s| = |m_d/m_e| = 3\eeq which are in
better agreement with the measured values. Now, under SO(10) $\to$
SU(5)$\,\times\,$U(1) the $\bf{126}$ contains precisely a
$\mathbf{45}$ which, as Harvey, Ramond and Reiss
showed~\cite{Harvey:1980je,Harvey:1981hk}, allows to implement the
same mechanism also in SO(10).

In our SO(10) model we assume that the SM $SU(2)_L$ Higgs doublet is a
combination of representations in the $\bf{10}$ and $\bf{126}$ of
SO(10).  While this allows to account for non-unification for the
down-quark and charged lepton masses of the two lightest families, it
still predicts an approximate quark-lepton symmetry, that is $m_e\sim
m_d$ are in any case connected by coefficients of order 1. As regards
the neutrino Dirac mass matrix, for definiteness we will stick to the
simpler relation in \eqn{prima}, which can coexist with \eqn{seconda}
for $m_{\mu,s}$, $m_{e,d}$ if the $u$-$\nu$ sector masses are
dominated by the $\mathbf{10}$ vevs. We also assume that, in the
diagonal basis for the down-quarks and charged leptons mass matrices,
the unitary rotation $V_L$ that diagonalizes the symmetric matrix
$m_D$ coincides with the Cabibbo-Kobayashi-Maskawa (CKM) rotation that
diagonalizes $m_u$.  Namely we assume as a working hypothesis:
\beq
m_D= m_u  \qquad {\rm and} \qquad V_L = V_{CKM}\,.
  \label{eq:terza}
\eeq
We stress at this point that our results do not depend in any crucial
way on the precise form of the quark-lepton relations, and the ansatz
\eqn{eq:terza} is adopted here only for the sake of
simplicity. However, the possibility of constructing a predictive
framework does depend on the fact that in SO(10) a precise relation
between $m_D$ and $m_u$ and $V_L$ and $V_{CKM}$ exists, which
naturally follows from fermions unification within a single
irreducible representation of the group. In fact, once the details of
the symmetry breaking pattern and of the fermion couplings to the
$\mathbf{10}$ and $\mathbf{126}$ are given, a quark-lepton mass
relation remains in any case fixed, and in particular a highly
hierarchical spectrum for the eigenvalues of $m_D$ is a
straightforward consequence of the SO(10) GUT framework (see also
Ref.~\cite{Akhmedov:2003dg}).  As regards the full $6 \times 6$ mass
matrix of the neutral sector, recalling that symmetric Yukawa matrices
imply that $m_D^T =m_D$, it can be written as
 \beq
 M = \left(
\begin{array}{cc}
0  & m_D \\
 m_D   &  M_R
\end{array}
\right)\,,
\label{Mgr}
\eeq
where $m_D$ and $M_R$ receive respectively contributions from the following vevs:
\beq
\label{eq:vevs}
m_D \sim \langle\mathbf{10} + \mathbf{126}
\rangle_{\Lambda_{EW}}\,, \qquad \qquad M_R \sim \langle \mathbf{126}
\rangle_{\Lambda_R}\,,
\eeq
where the $\mathbf{126}$ contributing to $M_R$ has a vev ${\cal
  O}(\Lambda_R)\gg \Lambda_{EW}$ along the $SU(5)$ singlet component.

\section{Compact RH neutrino spectrum}
\label{motiv}

The hierarchy $\Lambda_{EW}/\Lambda_R\ll 1$ between the two types of
vevs in \eqn{eq:vevs} enforces the see-saw mechanism, and,
after diagonalizing the matrix (\ref{Mgr}), one obtains the light
neutrino mass matrix $m_\nu$ from the seesaw formula that, with
$m_{D}^{T}=m_D$, reads:
\begin{eqnarray}
m_{\nu} & \simeq  & -m_{D}\, M_{R}^{-1}\,m_{D}\,.
\label{seesaw}
\end{eqnarray}
Inverting the seesaw formula (\ref{seesaw}) gives
\beq
M_R \simeq - m_D\, m_\nu^{-1}\, m_D \label{MR-1},
\eeq
which shows that one can obtain information on $M_R$ by using the
available experimental data on $m_\nu$, and assuming quark-lepton
symmetry for $m_D$.

Quark-lepton symmetry however, renders problematic the implementation
of the mechanism of baryogenesis via leptogenesis within the SO(10)
see-saw~\cite{Falcone:2000ib, Nezri:2000pb}. This is due to the two
factors of $m_D$ in \eqn{MR-1} that in general yield a very
hierarchical spectrum for the RH neutrinos. In fact, by fixing the
intermediate scale $\Lambda_R$ around $10^{11}\,$GeV, the lightest RH
state $N_1$, which is generally the main one responsible for
generating a lepton asymmetry, acquires a mass $M_{R_1}\ll
10^{9}\,$GeV, that is well below the Davidson-Ibarra (DI)
limit~\cite{di} which gives the benchmark to guarantee a sufficient
production of lepton asymmetry from RH neutrino decays.  There are
basically two ways out to this problem. The first one relies on the
fact that under certain conditions leptogenesis can also proceed via
the decays of the two heavier RH neutrinos~\cite{N2}, whose masses
remain well above the DI bound.
Refs.~\cite{oscar,db1,Abada:2008gs,DiBari:2008mp,DiBari:2010ux,Blanchet:2010td}
present specific realizations of this possibility.  The second way out
relies on the possibility of enhancing resonantly the CP
asymmetries~\cite{PU}, which allows to evade completely the DI bound,
but requires that at least one pair of RH neutrinos is highly
degenerate in mass.
In this paper we explore a third possibility,   namely that in spite of the
quark-lepton symmetry, the RH neutrino spectrum could still turn out
to be of a compact form~\cite{Buccella:2010jc} that is, characterized by at most mildly
hierarchical mass eigenvalues, all with values within the range
$10^{11\pm 2}\,$GeV which is the optimal one for leptogenesis.
Clearly, such a possibility would avoid from the start the problem of
a too light $N_{1}$.  Let us see in detail how this possibility can be
implemented.

\subsection{Conditions for a compact $N_i$ spectrum}
\label{compact}

A generic Dirac neutrino mass $m_D$ can be diagonalized by means of a
biunitary transformation with two unitary matrices $V_L$ and $V_R$,
that is $ m_D= V_L^\dagger m_D^{diag} V_R \label{MD-1} $.  However,
due to the assumed specific symmetry breaking pattern, in our model
$m_D$ is {\it symmetric}, and in this case there exists (Takagi
factorization \cite{Takagi}) a single unitary matrix $V_L$ such that
\begin{eqnarray}
m_{D} & = & V_{L}^{\dagger}m_{D}^{diag}V_{L}^{*},
\end{eqnarray}
where $m_{D}^{diag}={\rm diag}\left(m_{D1},m_{D2},m_{D3}\right)$ is
diagonal with real and non-negative eigenvalues.
It follows that the RH neutrino mass matrix can be written in the form
\begin{eqnarray}
M_{R} & = &
-V_{L}^{\dagger}m_{D}^{diag}A \, m_{D}^{diag}V_{L}^{*}  \,,
\label{eq:MRAL}
\end{eqnarray}
where we have introduced the symmetric matrix
\beq
A= V_L^\star\, m_\nu^{-1}\, V_L^\dagger\,.
\label{eq:A}
\eeq
On a naturalness ground, from the current knowledge about the light
neutrino mass matrix $m_{\nu}$, in the basis where charged lepton mass
matrix is diagonal one would expect that the elements of $A$ are at
most mildly hierarchical.  Then, if $m_D^{diag}$ is hierarchical as
implied by quark-lepton symmetry, we would generally obtain a
hierarchical RH neutrino spectrum.
Therefore only a quite specific structure of the $A$ matrix in
\eqn{eq:A} can enforce the conditions that ensure that the RH neutrino
spectrum is compact. To illustrate this issue, let us recall that we
are working under the assumption of quark-lepton
symmetry~\eqn{eq:terza} which implies in particular that in the basis
where the mass matrix of the charged leptons and of the down-type
quarks are diagonal $V_L = V_{CKM}$. Although the results in
Sections \ref{low} and \ref{LEPTO} are obtained with the assumption
$V_L = V_{CKM}$, to write down reasonable analytical expressions for
the RH neutrino mass spectrum in this section, we will set in first
approximation $V_{L}=I_{3\times3}$ (where the $ I_{3\times3}$ is the
$3\times 3$ identity matrix).  \Eqn{eq:MRAL} then gives
\begin{eqnarray}
\label{MRapprox}
M_{R} & \approx &
-\left(\begin{array}{ccc}
A_{11}m_{D1}^{2} & A_{12}m_{D1}m_{D2} & A_{13}m_{D1}m_{D3}\\
A_{12}m_{D1}m_{D2} & A_{22}m_{D2}^{2} & A_{23}m_{D2}m_{D3}\\
A_{13}m_{D1}m_{D3} & A_{23}m_{D2}m_{D3} & A_{33}m_{D3}^{2}
\end{array}\right)\,,
\end{eqnarray}
with $A\approx m_{\nu}^{-1}$ (recall that because $A$ is symmetric
$A_{ij} = A_{ji}$).  Quark-lepton symmetry implies $m_{D3}\gg
m_{D2}\gg m_{D1}$, which suggests that a generically compact RH
spectrum would result if
\begin{equation}
  \label{eq:compact}
\left|\frac{A_{33}}{A_{22}} \right| \lsim
\frac{m_{D2}^2}{m^2_{D3}} \qquad \qquad {\rm and}\qquad \qquad
\left|\frac{A_{23}}{A_{22}} \right|
\lsim \frac{m_{D2}}{m_{D3}}
\end{equation}
 since, if this were the case, all the
hierarchically large entries in $M_R$ would be sufficiently suppressed
(notice that the $M_R$(1,3) and $M_R$(2,2) elements of \eqn{MRapprox}
are non-hierarchical because from the light neutrino mass matrix we
expect $A_{22} \sim A_{13}$ and from the quark-lepton symmetry we
expect $m_{D1} m_{D3} \sim m_{D2}^2$).  However, in this paper we will
assume the more restrictive condition:
\begin{equation}
  \label{eq:hierarchy}
\left|\frac{A_{33}}{A_{11}} \right| \lsim \frac{m_{D1}^2}{m^2_{D3}}\ll 1, \qquad\qquad
\left| \frac{A_{23}}{A_{11}} \right|\lsim \frac{m_{D1}^2}{m_{D2}m_{D3}}\ll 1,\qquad\qquad
\end{equation}
since they are needed to justify the simplifying approximation
in~\eqn{eq:conditions} below.  The interest in exploring a scenario in
which the two conditions in \eqn{eq:hierarchy} are realized stems from
the fact that it is quite likely that an $A$-matrix of this form would
render leptogenesis a viable mechanism to explain the BAU within the
SO(10) seesaw framework.
In this paper, we will not speculate on the possible origin of the two
relations in~\eqn{eq:hierarchy}, nor we will attempt to reproduce them
by starting from a suitable fundamental Lagrangian, and thus the
possibility that such a pattern could arise basically relies only on
the fact that similar hierarchies do exist among quantities related to
the Yukawa coupling sector.
In fact, we believe that building up  a theoretical justification
for $A_{33}/A_{11},\, A_{23}/A_{11} \ll 1$  could be equally difficult than explaining
the mass hierarchies of the charged fermions, which is a long standing unsolved problem
in particle physics.  We will, however,  prove that as long as
$\theta_{13}$ is nonvanishing, imposing such relations is a
technically consistent procedure, in the sense that they can be always
fulfilled, regardless of the specific types of quark-lepton symmetry
relations assumed.  If $A_{23}$ and $A_{33}$ are negligible with
respect to all the other entries, we can set in first approximation
\begin{equation}
  \label{eq:conditions}
 A_{23} =A_{33}=0\,.
\end{equation}
As we will see, from these two conditions it follows that, besides
obtaining a compact RH spectrum, two eigenvalues in $M_R$ will
actually be
close to
degenerate.  In general, the degeneracy of pairs of $M_R$ eigenvalues
represents an interesting situation for leptogenesis, since it can
allow for resonant enhancement of the CP asymmetries.
Although we will find that, eventually, conditions \eqn{eq:conditions}
are not sufficient to bring the dynamics of leptogensis fully within
the resonant regime, it is still worth studying which class of general
conditions for $A$ could yield a pair of eigenvalues very close in
mass\footnote{The fact that our results for the Cosmic baryon
  asymmetry do not benefit from resonant enhancements of the CP
  asymmetries justifies the claim that the compact RH spectrum
  scenario represents a  third possibility for realizing
  leptogenesis within the SO(10) GUT.}.

The eigenvalues $\lambda$ of $M_{R}$ in \eqn{MRapprox} are given by
the solutions to the characteristic cubic equation
\begin{eqnarray}
\lambda^{3}+b\lambda^{2}+c\lambda+d & = & 0,
\label{eq:char_eq}
\end{eqnarray}
with
\begin{eqnarray}
b & = & A_{11}m_{D1}^{2}+A_{22}m_{D2}^{2}+A_{33}m_{D3}^{2},\nonumber \\
\nonumber
c & = & \left(A_{11}A_{22}-A_{12}^{2}\right)m_{D1}^{2}m_{D2}^{2}
+\left(A_{11}A_{33}-A_{13}^{2}\right)m_{D1}^{2}m_{D3}^{2}
+\left(A_{22}A_{33}-A_{23}^{2}\right)m_{D2}^{2}m_{D3}^{2},\\
d & = & \left(2A_{12}A_{13}A_{23}+A_{11}A_{22}A_{33} -A_{13}^{2}A_{22}-A_{12}^{2}A_{33}-A_{23}^{2}A_{11}
\right)m_{D1}^{2}m_{D2}^{2}m_{D3}^{2}.
\end{eqnarray}
The necessary condition for two eigenvalues being equal is that
the discriminant of \eqn{eq:char_eq} vanishes. We can write down the discriminant as follows
\begin{eqnarray}
\Delta & = & b^2 c^2 - 4 c^3 - 4 b^3 d + 18 b c d - 27 d^2 \nonumber \\
& \approx & \left(A_{23}^{2}-A_{22}A_{33}\right)^{2} m_{D2}^{4}m_{D3}^{4}
\left[A_{22}^{2}m_{D2}^{4}+A_{33}^{2}m_{D3}^{4}+2\left(2A_{23}^{2}
-A_{22}A_{33}\right)m_{D2}^{2}m_{D3}^{2}\right]\,,
\label{eq:discri}
\end{eqnarray}
where in the second line we have expanded
up to first order in $\frac{m_{D1}}{m_{D3}}\sim
\frac{m_{u}}{m_{t}}$. We have $\Delta=0$   if
\begin{eqnarray}
A_{23}^{2} & = & A_{22}A_{33},\;\;\;\;{\rm or}\;\;\;\; A_{23}^{2}
=-\left(\frac{A_{22}m_{D2}^{2}-A_{33}m_{D3}^{2}}{2m_{D2}m_{D3}}\right)^{2}.
\end{eqnarray}
We will consider only the first possibility, that involves solely
elements of the matrix $A$.  We then see that if
$A_{23}\,,A_{33}\approx0$ or alternatively $A_{23},\,A_{22}\approx0$,
quasi degeneracy of two RH neutrino masses results.  Notice that the
first condition also satisfies \eqn{eq:hierarchy}, and then it will
result in a compact spectrum, in contrast, as we will see in the
following, the second condition will yield a hierarchical spectrum.
Before dealing with these two cases in detail, let us remark that
without loss of generality, it is convenient to work in the basis
where the RH neutrino mass matrix $M_R$ is diagonal.  Since $M_{R}$ is
symmetric, it can be brought to diagonal form $M_{R}^{diag}={\rm
  diag}(M_1,M_2,M_3)$ with real and positive entries by means of a
unitary matrix $W$:
\begin{eqnarray}
M_{R}^{diag} & = & W^{\dagger}M_{R}W^{*}\,.
\label{eq:MR_d}
\end{eqnarray}
In this basis we redefine the Dirac mass matrix as follows
\begin{eqnarray}
\hat{m}_{D} & = & m_{D}W^{*} \label{eq:mdtilde}.
\end{eqnarray}
From now on we will always work in this basis.

\subsubsection{Case 1: $A_{23}=0$, $A_{33}=0$}

In this case, we solve \eqn{eq:char_eq} and expand the eigenvalues up to first order in
$\frac{m_{D1}}{m_{D3}}$; we obtain the following spectrum for the RH
neutrinos\footnote{Notice that the physical RH neutrino masses, eqs. (\ref{Mrval}),
  correspond to the absolute value of eigenvalues obtained from
  solving \eqn{eq:char_eq}.  Alternatively, one can also find the
  unitary matrix $W$ in ~\eqn{eq:MR_d} which diagonalizes $M_R$ up
  to first order in $\frac{m_{D1}}{m_{D3}}$.}
\begin{eqnarray}
M_{1} & = & \left|A_{22}\right|m_{D2}^{2},\nonumber \\
M_{2} & = & \left|A_{13}\right|m_{D1}m_{D3},\nonumber \\
M_{3} & = & \left|A_{13}\right|m_{D1}m_{D3} \label{Mrval}.
\end{eqnarray}
With the reasonable assumption that $\left|A_{13}\right|$ and $\left|A_{22}\right|$ are not
very hierarchical, we see that it is possible to have $M_{1}\simeq
M_{2,3}$ and, depending on the values of $m_{D}^{diag}$ renormalized
at the leptogenesis scale, both mass orderings $M_{1}<M_{2,3}$ or
$M_{2,3}<M_{1}$ are possible.
In ref.~\cite{Buccella:2010jc} only the $M_{1}<M_{2,3}$ ordering was
considered, a vanishing $\theta_{13}$ was assumed, and both lepton flavour and
heavier RH neutrino effects in leptogenesis had been  ignored.
Instead, as we will show, the ordering $M_{2,3}<M_{1}$ can indeed
occur, $\theta_{13}\neq 0$ is a crucial condition to ensure the
existence of solutions for the compact spectrum conditions, and as
regards the lepton flavour and heavy RH neutrino effects, they must be
included in order to obtain successful leptogenesis, and to guarantee
that the result is reliable.

\subsubsection{Case 2: $A_{23}=0$, $A_{22}=0$}

In this case, by proceeding as before up to first order in
$\frac{m_{D1}}{m_{D3}}$, we obtain the spectrum
\begin{eqnarray}
M_{1} & = & \left|A_{33}\right|m_{D3}^{2},\nonumber \\
M_{2} & = & \left|A_{12}\right|m_{D1}m_{D2},\nonumber \\
M_{3} & = & \left|A_{12}\right|m_{D1}m_{D2}.
\end{eqnarray}
Assuming also in this case that $\left|A_{12}\right|$ and
$\left|A_{33}\right|$ are not exceedingly hierarchical implies $ M_{1}
\gg M_{2,3}$. Of course in this case, since the large contributions
from $m_{D3}$ are not suppressed, we do not expect to obtain a compact
RH spectrum. Nevertheless, in principle leptogenesis could still
proceed at a scale $M_{2,3}\ll \Lambda_R$ thanks to
the asymmetries generated in the decays of the two quasi degenerate
states $N_{2,3}$.
Eventually however, we will find that in Case 2 on the one hand
leptogenesis is unable to produce a sufficient baryon asymmetry, and
on the other hand for the heaviest RH neutrino we always obtain
$M_3\gsim 10^{14}\,$GeV which, under the requirement of perturbative
Yukawa couplings, is in conflict with $SO(10)$ gauge coupling
unification which instead suggests  an intermediate vevs scale
of order $10^{11}\,$GeV~\cite{Bertolini:2009qj}
(see however~\cite{Bertolini:2012im} for viable scenarios
with an intermediate scale as high as $10^{14}$ GeV).

\section{Relation with low energy observables}
\label{low}

We have seen that by assuming conditions~\eqn{eq:conditions} we have
forcibly ended up with a quasi degenerate pair of RH eigenvalues.
Before proceeding, let us stress that while setting the values of
$A_{23}$ and $A_{23}$ to an exact zero has the virtue of simplifying
the analysis, a generic compact RH neutrino spectrum can be obtained
by fixing instead their values to any sufficiently small number as
dictated by \eqn{eq:compact}.  Doing this would lift the quasi
degeneracy, but would still yield similar results.  We will return to
this point in Section~\ref{LEPTO}.
One important point is that
requiring that the matrix $A$ satisfies some specific conditions gets
reflected in specific relations between the low energy observables,
and yields an enhanced level of predictability for the SO(10) model.
Before studying which type of relations arise, it is useful to carry
out a quick counting of the fundamental free parameters of the theory,
and list the phenomenological constraints that they should satisfy.
The structure of the two symmetric matrices $m_D$ and $M_R$ is
determined by two corresponding sets of fundamental Yukawa couplings
between the fermion fields in the $\mathbf{16}$ and the Higgs fields
respectively with vevs $\sim \Lambda_{EW}$ and $\sim
\Lambda_{R}$. This amounts to $6+6+1=13$ real parameters corresponding
to the two symmetric Yukawa matrices plus the ratio $\Lambda_{EW}/
\Lambda_{R}$ which determines the seesaw suppression of neutrino
masses, or in other words their absolute scale.  Under the assumption
of quark-lepton duality (see eq. \ref{eq:terza}), the values of the 13
real parameters are constrained by the following observables in the
up-type quark and neutrino sectors: the three quark masses
$m_u,\,m_c,\,m_t$, the two neutrino mass-squared differences $\Delta
m^2_{12},\, \Delta m^2_{23}$, the three CKM mixing angles
$\theta_{12}',\,\theta_{23}',\,\theta_{13}'$ and the three PMNS mixing
angles $\theta_{12},\,\theta_{23},\,\theta_{13}$, which add up to a
total of $11$ constraints.  Now, imposing on the complex elements of
the matrix $A$ two additional conditions, e.g. $A_{23}=A_{33}=0$
(or any other pair of conditions), implies that the set of 13 real
fundamental parameters must satisfy two additional requirements, that
in our case read ${\rm Re}(A_{23})={\rm Re}(A_{33})=0$.
Thus the parameter space of the model remains completely determined
allowing to obtain a quantitative prediction for the absolute neutrino
mass scale $m_{\nu_1}$.  As regards the constraints on imaginary
quantities, there are many fundamental complex phases, and only one
measured observable, the CKM phase $\delta'$. Nevertheless, as we will
see, the structure of the conditions
implies nontrivial relations between $\delta'$ and the three PMNS phases
$\alpha,\beta$ and $\delta$.

In the following we assume a hierarchical and normally ordered
spectrum for the light neutrino masses
 $ m_\nu^{diag} = {\rm{diag}} (m_{1}, m_{2}, m_{3}) $
with
\beq
m_{1} < m_{2} < m_{3},
\eeq
which is justified by the assumption of quark-lepton
symmetry~\eqn{eq:terza}. In the basis where charged lepton mass matrix is diagonal,
the PMNS  mixing matrix $U_{PMNS}$
diagonalizes the effective neutrino mass matrix
\beq
m_\nu = U_{PMNS}^\star m_\nu^{diag} U_{PMNS}^\dagger.  \label{MLL}
\eeq
We adopt for $U_{PMNS}$ the standard parametrization in terms of 3
angles and three complex phases:
%
\begin{eqnarray}
\label{eq:UPMNS}
U_{PMNS} & = & \left(\begin{array}{ccc}
c_{12}c_{13} & s_{12}c_{13} & s_{13}e^{-i\delta}\\
-s_{12}c_{23}-c_{12}s_{23}s_{13}e^{i\delta} & c_{12}c_{23}-s_{12}s_{23}s_{13}e^{i\delta} & s_{23}c_{13}\\
s_{12}s_{23}-c_{12}c_{23}s_{13}e^{i\delta} & -c_{12}s_{23}-s_{12}c_{23}s_{13}e^{i\delta} & c_{23}c_{13}\end{array}\right)
\times{\rm diag}\left(1, e^{i\alpha}, e^{i\beta} \right)\,.
\end{eqnarray}
Here $c_{ij}=\cos \theta_{ij}$ and $s_{ij}=\sin \theta_{ij}$, with $i$
and $j$ labeling families that are coupled through that angle ($i, j =
1, 2, 3$).  Note that since the computation of the leptogenesis CP
asymmetries involves several interfering amplitudes, the angles
$\theta_{ij}$ cannot be restricted to the first quadrant, except for
$\theta_{13}$ that can be taken to be positive once the CP phase
$\delta$ is allowed to range between $-\pi$ and $\pi$.

According to the quark-lepton symmetry ansatz~\eqn{eq:terza}, in the numerical
analysis we take $V_L=V_{CKM}$, and accordingly we parametrize
$V_L$ with three angles and one phase, with a structure analogous to
the first matrix on the right-hand-side (RHS) of \eqn{eq:UPMNS},
distinguishing the angles and phase with a  prime superscript:
$s_{12}',\,s_{23}',\,s_{13}',\,\delta'$\footnote{ Taking $V_L\approx
  V_{CKM}$ implies that the large leptonic mixing observed in the
  low-energy sector should be a consequence of a see-saw enhancement
  of lepton mixing. Such an enhancement requires a strong (quadratic)
  mass hierarchy of RH neutrinos, or a $M_R$ structure with large
  off-diagonal entries~\cite{Smirnov:1993af}.}.  The following
discussion, however, is based on analytical expressions that get
largely simplified by writing $V_L$ in the approximate Cabibbo-like
form:
\beq V_L^{(C)} = \left(
\begin{array}{ccc}
 \cos \theta_C & \sin \theta_C & 0 \\
 -\sin \theta_C & \cos \theta_C & 0 \\
 0 & 0 & 1
\end{array}
\right)\,.
\label{CKMtype}
\eeq
We will then write our formulae in this approximation, keeping in mind
however, that the full expressions have been used to obtain the
numerical results.

The matrix $A$ \eqn{eq:A} can  be expressed in terms
of the observables  $V_L$,  $U_{PMNS}$ and  $ m_\nu^{diag}$ as
\begin{equation}
  \label{eq:AVUm}
  A= \left(V_L U_{PMNS}^\star\right)^\star\, \frac{1}{m_\nu^{diag}}\,
\left(V_L  U_{PMNS}^\star\right)^\dagger\,.
\end{equation}
In the approximation $V_L=V_L^{(C)}$, the conditions ${A}_{23}=
{A}_{33}=0 $ yield the following two relations:
\begin{eqnarray}
\label{eq:cond12}
  \frac{m_2}{m_1}\, e^{-2 i \alpha } &=&
-\frac{\left(
c_{12} s_{23}+e^{i \delta} s_{12} s_{13} c_{23}\right)
\left[s_c c_{12} s_{13} s_{23}+e^{i \delta }
(s_c s_{12} c_{23}- c_c c_{12}
      c_{13})\right]}
{\left(-s_{12} s_{23}+e^{i \delta } c_{12}
      s_{13} c_{23}\right) \left[- s_c s_{12} s_{13}
s_{23}+e^{i \delta } (c_c s_{12} c_{13}+ s_c c_{12}
      c_{23})\right]} \,,
 \\
\label{eq:cond13}
  \frac{m_3}{m_1}\, e^{-2 i \beta}&=&
  \frac{c_{13} c_{23} \left[s_c s_{12} c_{23}+c_{12}
\left(- c_c c_{13}+ s_c e^{-i \delta }
s_{13} s_{23}\right)\right]}
{\left(-s_{12} s_{23}+e^{i \delta}
c_{12} s_{13} c_{23}\right)
\left(s_c c_{13} s_{23}+ c_c e^{i \delta } s_{13}\right)}.
\end{eqnarray}
where $s_c,c_c= \sin\theta_C,\cos\theta_C$.

We see that by taking the absolute values of
\eqns{eq:cond12}{eq:cond13} we obtain two conditions that do not
depend on the Majorana phases $\alpha,\beta$ (and are also  even functions
of $\delta$ that depend only on its cosine).  We can further eliminate
$m_2$ and $m_3$ by using their relations with the solar and
atmospheric mass-squared differences\footnote{The relation with the
  atmospheric mass-squared difference $m^2_3-c^2_s m^2_2-s^2_s m^2_1 =
  \Delta m^2_a$ (see ref.~\cite{Buccella:2010jc}) can be equally well
  used with irrelevant numerical differences.}:  $m^2_2=m^2_1+\Delta
m^2_s$ and $m^2_3=m^2_1+\Delta m^2_a$, obtaining:

\begin{eqnarray}
  \label{eq:delta1}
1+\frac{\Delta m^2_s}{m_{1}^2} &=&
f_{12}([\theta_C,\theta_{12},\theta_{23},\theta_{13}];\cos\delta),\\
  \label{eq:delta2}
1+\frac{\Delta m^2_a}{m_{1}^2} &=&
f_{23}([
\theta_C,\theta_{12},\theta_{23},\theta_{13}];\cos\delta) \,.
\end{eqnarray}
The absolute neutrino mass scale $m_{1}$ appearing on the LHS of
these equations represents the first unknown.  On the RHS, $f_{12}$ and
$f_{23}$ are two known (although non-transparent) functions of known
mixing angles (that are listed within the squared brackets) and of the
cosine of the second unknown, that is the Dirac phase $\delta$.  These
two equations might or might not have physical solutions (for example,
given that $\Delta m^2_s >0$, in case the RHS of the first equation
remains $\leq 1$ for all values of $\delta$, there are no physically
acceptable solutions). If solutions exists, these will corresponds to
specific values of $m_{1}$ and of $\delta$ with
uncertainties determined by the experimental errors on the mixing
angles.

As regards the conditions on the complex arguments of
\eqns{eq:cond12}{eq:cond13}, they have the form
\begin{eqnarray}
  \label{eq:alpha}
\alpha =
g_{12}([\theta_C,\theta_{12},\theta_{23},\theta_{13},\Delta m_s^2];\delta,m_1),\\
  \label{eq:beta}
\beta =
g_{23}([\theta_C,\theta_{12},\theta_{23},\theta_{13},\Delta m_a^2];\delta,m_1),
\end{eqnarray}
%
%
%
where $g_{12}$ and $g_{13}$ are again known functions, and thus
$\alpha$ and $\beta$ can be determined in terms of $m_1$, $\delta$ and
of the known mixing angles and mass squared differences.
This completes the determination of all
the low energy observables in terms of measured quantities, and
through \eqn{eq:MRAL} also fixes the RH neutrinos mass spectrum.

\begin{table}[t!]
\begin{center}
\begin{tabular}{|c|c||c|c|}
  \hline
  \multicolumn{2}{|c||}{\rm Quark\ sector } &
  \multicolumn{2}{|c|}{\rm Neutrino\ sector }  \\ \hline
  &&& \\ [-8pt]
  $m_u(\Lambda)$ &$ 0.00067\> {\rm GeV}$&$ \Delta m^2_{21}(\Lambda)$&$ 11.86 \times 10^{-5}\> {\rm eV}^2$ \\  [2pt]
  $m_c(\Lambda)$ &$ 0.327\ \  {\rm GeV}$&$\Delta m^2_{31}(\Lambda)$&$ 3.84 \times 10^{-3}\> {\rm eV}^2$ \\  [2pt]
  $m_t(\Lambda)$ &$ 99.1\> \ \ \ \; {\rm GeV} $&   & \\  [4pt]
  \hline &&& \\ [-8pt]
  $\theta'_{12}$ &$ 13.02^\circ$ &$ \theta_{12}  $ &$ 34.4^\circ $\\  [2pt]
  $\theta'_{23}$ &$ \ \, 2.35^\circ $&$ \theta_{23}  $ &$ 42.8^\circ $\\  [2pt]
  $\theta'_{13}$ &$ \ \, 0.20^\circ$ &$ \theta_{13}  $ &$ \ 5.6^\circ$ \\  [2pt]
  $\delta'$     &$\quad\ \;  1.20\,$rad&    &  \\ [4pt]
  \hline
\end{tabular}
\caption{Input parameters. We use the up-quark masses renormalized to the
  scale $\Lambda = 10^{9}$ given in Table IV in
  Ref.~\cite{Xing:2007fb}. Neutrinos mass squared differences are
  taken from the global fit in Ref.~\cite{GonzalezGarcia:2010er} and
  renormalized to the scale $\Lambda$ with a multiplicative factor
  $r^2$ with $r=1.25$ according to the prescription in
  Ref.~\cite{thermal}.  The CKM mixing angles $\theta'_{ij}$ and CKM
  phase $\delta'$ are derived from the values of the Wolfenstein
  parameters given in Ref.~\cite{PDG10}.  The PMNS mixing angles are
  taken from the global fit in
  Ref.~\cite{GonzalezGarcia:2010er}. Renormalization effects for the
  CKM and PMNS parameters have been neglected.}
\label{tab:1}
\end{center}
\vspace{-0.9cm}
\end{table}

To be more precise, given that the signs of $\theta_{12},\theta_{23}$
and $\theta_{13}$ are not determined in oscillation experiments,
depending on the possible choices $\pm\theta_{ij}$ the two
\eqns{eq:delta1}{eq:delta2} represent in principle $2^3=8$ conditions.
However, the PMNS phase always appears together with $\theta_{13}$ in
the combination $s_{13}e^{i\delta}= -s_{13}e^{i(\delta\pm \pi)}$ so
that if $\delta$ is a solution for $+\theta_{13}$, $\delta\pm \pi$ is
a physically indistinguishable solution for $-\theta_{13}$, and this
reduces the eight possible pairs of equations to just four.

\begin{figure}[t!!]
\begin{center}
\vspace{1cm}
 \includegraphics[width=7.4cm,height=6cm,angle=0]{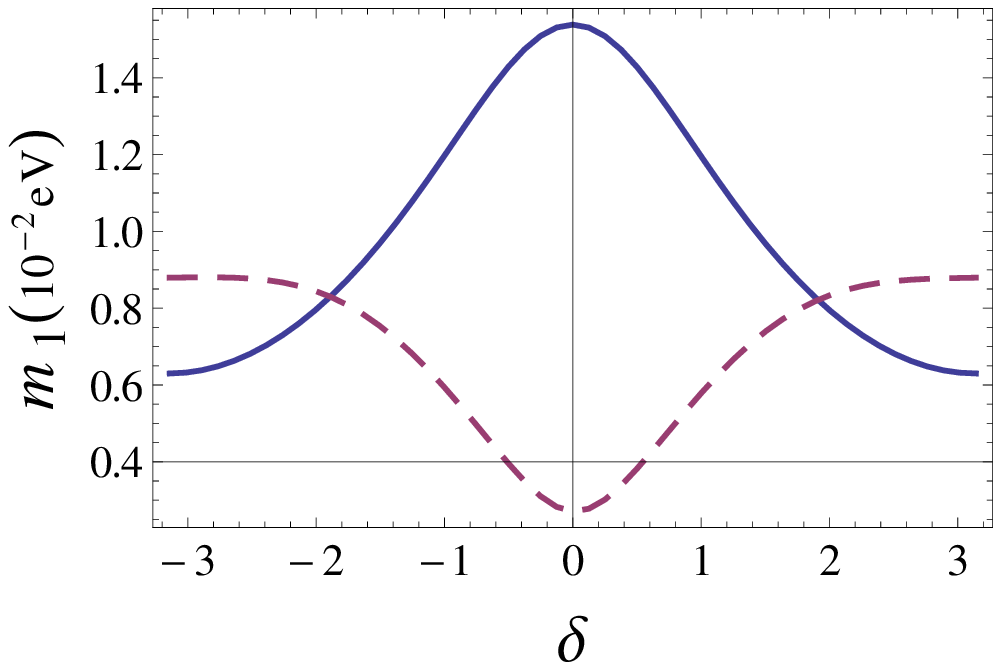}
\hspace{.5cm}
 \includegraphics[width=7.4cm,height=6cm,angle=0]{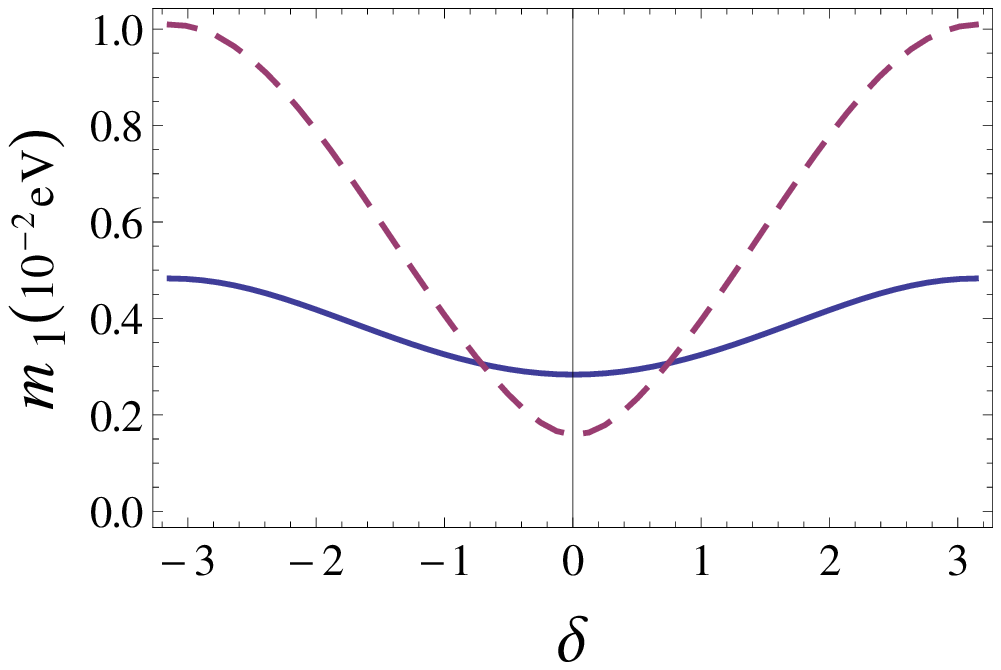}
\caption{Plots of $m_{1}$ as a function of $\delta$ according to
  \eqn{eq:delta1} (solid blue line) and \eqn{eq:delta2} (dashed violet
  line). The points of intersection represent pairs of possible
  solutions for $(m_{\nu_1},\delta)$.  Left panel
  $(+|\theta_{12}|,-|\theta_{23}|)$,
right panel
  $(-|\theta_{12}|,-|\theta_{23}|)$, with
  $(|\theta_{12}|,|\theta_{23}|)=(34.4^\circ,42.8^\circ)$.
  \label{fig:1}}
\end{center}
\end{figure}

Once the simplification $V_L \to V_L^{(C)}$ is dropped, due to the
presence of the CKM phase $\delta'$ \eqns{eq:delta1}{eq:delta2}
acquire a (mild) dependence also on $\sin\delta$, meaning that for
each one of the four possibilities $(\pm \theta_{12},\pm \theta_{23})$
we can have two nonequivalent solutions corresponding to values of
$\delta$ of opposite signs. An example of this situation is
illustrated in fig.\ref{fig:1} for the two cases
$(+,-)\equiv(+|\theta_{12}|,-|\theta_{23}|)$ and
$(-,-)\equiv(-|\theta_{12}|,-|\theta_{23}|)$.  The two relations
\eqn{eq:delta1} and \eqn{eq:delta2} correspond to two different curves
$m_{1}(\delta)$ that are plotted respectively with the solid blue
lines and the dashed violet lines and intersect in  two points
that are the solutions to the system of constraints.
Notice that a solution to these constraints does not always exists i.e. when
the two curves do not intersect. This happens for example in some of the scenarios in
Case 2, as shown in Table~\ref{tab:3}, that has therefore
less entries than Table~\ref{tab:2} of Case 1.

\begin{figure}[t!]
\begin{center}
 \includegraphics[width=7.4cm,height=6cm,angle=0]{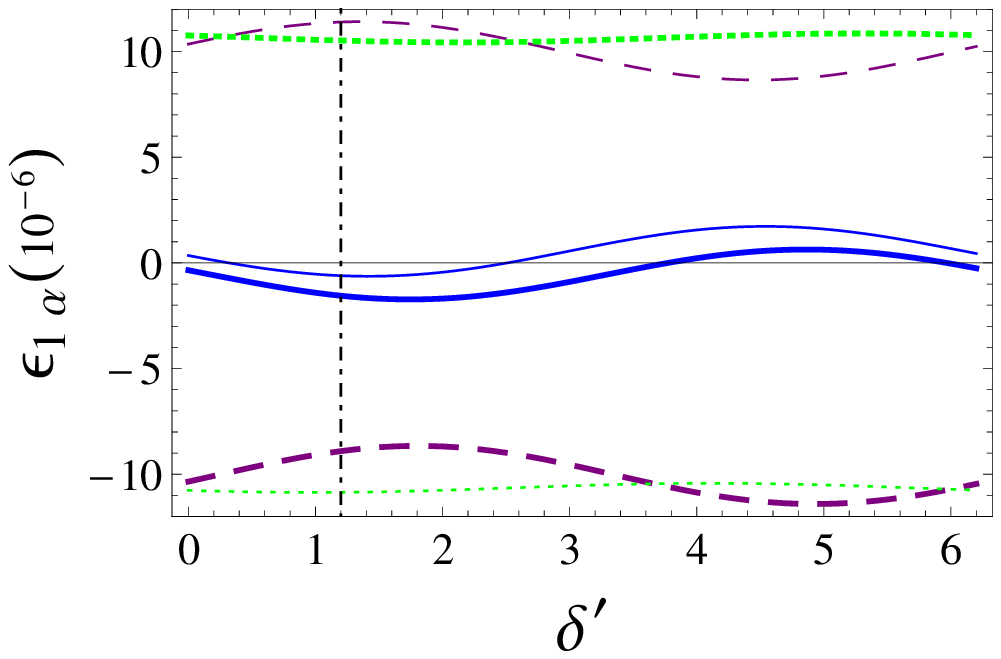}
\hspace{.5cm}
 \includegraphics[width=7.4cm,height=6cm,angle=0]{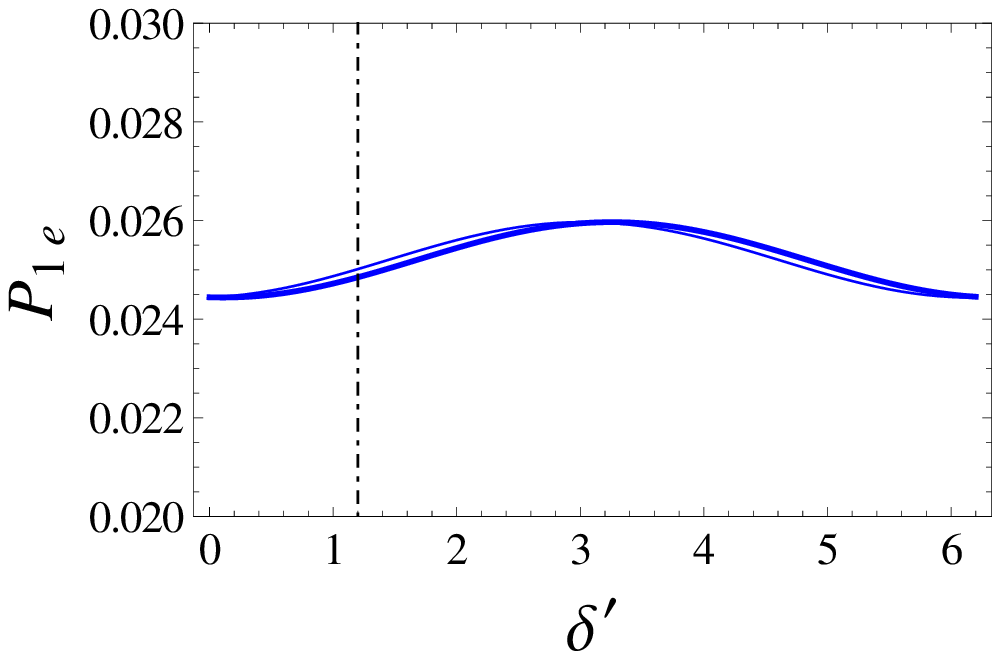}
\caption{Left: the $N_1$  CP asymmetries
$\epsilon_{1e}$ (solid blue lines)
$\epsilon_{1\mu}$ (dashed violet lines)
$\epsilon_{1\tau}$ (dotted green lines)
for the solution labeled $(-,-)$ in Table~\ref{tab:2} (Case 1)
for positive (thick lines) and negative (thin lines) values
of the Dirac phase $\delta$, as a function of the $V_L$ phase
$\delta'$. Right: the $N_1$ electron flavour washout projector
$P_{1e}$    for positive (thick line) and negative (thin line) values
of  $\delta$ as a function of $\delta'$. The vertical dot-dashed lines correspond to
the value $\delta'=\delta_{CKM}=1.20\,$rad
used in the numerical analysis.
\label{fig:2}}
\end{center}
\end{figure}

The input parameters of our numerical analysis are listed in
Table~\ref{tab:1}.  For the eigenvalues of $m_D$ we use the values of
the up-quark masses renormalized to the scale $\Lambda  = 10^{9} $ GeV ($\sim M_R$),  given
in Table IV in Ref.~\cite{Xing:2007fb}. The relevant values of $M_R$ we find are given in Table~\ref{tab:2}. Neutrinos mass square
differences are taken from the global fit in
Ref.~\cite{GonzalezGarcia:2010er} and renormalized to the scale
$\Lambda$ with a multiplicative factor $r^2$ with $r=1.25$
according to the prescription in Ref.~\cite{thermal}.  The CKM mixing
angles $\theta'_{ij}$ and CKM phase $\delta'$ are derived from the
values of the Wolfenstein parameters given in Ref.~\cite{PDG10},
renormalization effects for these angles are small and have been
neglected.

As regards the PMNS mixing angles, recent fits to oscillation neutrino
data suggest a small but nonvanishing value for $\theta_{13}$.  In our
scenario, having $\theta_{13} \neq 0$ is of fundamental importance
because only under this condition the Dirac phase $\delta$ will enter
the constraining equations \eqn{eq:cond12} and \eqn{eq:cond13},
providing enough free parameters to allow for a numerical solution, so
let us discuss this specific quantity a bit more in detail.  With the
assumption of normal ordering $m_1 < m_2 < m_3$, and with $3\sigma$
errors for the three-flavour neutrino oscillation parameters, the
following results have been reported:
\beq
\sin^2 \theta_{13} =
0.013^{+0.023}_{-0.015} \text{\cite{Schwetz:2011zk}}, \qquad
\sin^2\theta_{13} = 0.025^{+0.025}_{-0.02} \text{\cite{Fogli:2011qn}},\qquad
\sin^2 \theta_{13} \leq 0.043 \text{\cite{GonzalezGarcia:2010er}}\,.
\label{theta13}
\eeq
The last result is a 3$\sigma$ upper limit estimated
in the framework of the so called GS98 solar model
with the Ga capture cross-section of Ref.~\cite{Bahcall:1997eg}.
At 1$\sigma$ the same data give $\sin^2 \theta_{13} =
0.0095^{+0.013}_{-0.007}$ corresponding to
$\theta_{13}=(5.6^{+3.0}_{-2.7})^\circ$ \cite{GonzalezGarcia:2010er}.
We use such best fit value, and also for the other two
angles we adopt the results of the global fit in
Ref.~\cite{GonzalezGarcia:2010er}.

Our results for the possible values of the still unmeasured low energy
parameters $m_{1},\,\delta,\,\alpha$ and $\beta$, and for the RH
neutrino masses evaluated according to \eqn{eq:MRAL}, are collected in
Table~\ref{tab:2} and in Table~\ref{tab:3}. Here onwards we always arrange the
ordering of RH neutrino masses according to $M_1 < M_2 < M_3$.
The first four lines of Table~\ref{tab:2} list the 4+4 possible
solutions for the conditions of Case 1 $A_{23}=A_{33}=0$.
In each line we list the two solutions
corresponding to positive and negative values of $\delta$. Note that
the numerical differences between the absolute values of each pair of
solutions for $\delta$ are small, since they correspond to effects
suppressed by $s_{13}'$. However, even such a small difference can
have a non negligible impact on the value of the leptogenesis CP asymmetries.
Real parameters like $m_{1}$ and
$M_i$ also come in pairs with very close values, but in this case the
differences are numerically irrelevant so that a single approximate
value is displayed.  This situation is illustrated in
Fig.~\ref{fig:2}: the left panel depicts the three $N_1$ flavoured CP
asymmetries (see the next section) $\epsilon_{1e}$ (solid blue lines)
$\epsilon_{1\mu}$ (dashed violet lines) $\epsilon_{1\tau}$ (dotted
green lines) for the solution labeled $(-,-)$ in Table~\ref{tab:2}
(Case 1) for positive (thick lines) and negative (thin lines) values
of the Dirac phase $\delta$, as a function of the $V_L$ phase
$\delta'$. We see that for the CP asymmetries which are very sensitive
to the values of the complex phases, the two different solutions for
$\delta$ induce very large effects, for example they swap completely
the signs of $\epsilon_{1\mu}$ and $\epsilon_{1\tau}$.  In the right
panel, as an example of one important real leptogenesis parameter, we
have plotted the electron flavour washout projector $P_{1e}$ for the
RH neutrino $N_1$ (see next section), for positive (thick line) and
negative (thin line) values of $\delta$ as a function of $\delta'$. We
see that in this case numerical differences are irrelevant.  In both
panels the vertical lines correspond to the value
$\delta'=\delta_{CKM}=1.20\,$rad that we have used in the numerical
analysis of Tables~\ref{tab:2}-\ref{tab:3}.

The last line  in Table~\ref{tab:2} labeled with $(-,-)^*$ gives
the results obtained for that case when $A_{23},\,A_{33}$ are set
to  small but nonvanishing  values, that we have  (arbitrarily)
chosen as (cfr. \eqn{eq:compact}):
\begin{equation}
  \label{eq:nonzero}
\left|\frac{A_{33}}{A_{22}} \right| = 0.05\times
\frac{m_{D2}^2}{m^2_{D3}} \qquad \qquad {\rm and}\qquad \qquad
\left|\frac{A_{23}}{A_{22}} \right|
= 0.05 \times \frac{m_{D2}}{m_{D3}}\,.
\end{equation}
We see that the changes in $\delta,\, m_1,\, \alpha$ and $\beta$ with
respect to the $A_{23}=A_{33}=0$ case in the fourth line remain below
the precision of the table, and in any case are way too small to be
seen experimentally. The same happens for all the other cases, and we
can thus conclude that the predictions obtained with our simplified
conditions~\eqn{eq:conditions} hold for each class of compact spectrum
solutions.  The last column of this line however, makes apparent that
the resulting RH spectrum is just compact rather than degenerate,
which justifies talking about a third way to SO(10) leptogenesis.

%
\begin{table}[t!]
\begin{center}
\renewcommand{\arraystretch}{1.3}
\begin{tabular}{|@{\;}c@{\;}|@{\;}c@{\;}|@{\;}c@{\;}
|@{\;}c@{\;}|@{\;}c@{\;}|@{\ }l@{\;}|}
\hline
$\left(\theta_{12},\theta_{23}\right)$ & $\delta$ &
$m_{1}\,$\small($10^{-3}$ eV) & $\alpha$ & $\beta$
&    $(M_1,\,M_2,\,M_3)\,$ \small($10^{9}\,$GeV)
\tabularnewline
\hline
\hline
$(+,+)$ & $(1.43,-1.46)$ & $9.7$ & $(-1.47,1.46)$ & $(-0.18,0.19)$ & $(3.5,\ 3.9,\ 3.9)$\tabularnewline
\hline
$(-,+)$ & $(2.60,-2.63)$ & $2.9$ & $(1.52,-1.51)$ & $(1.25,-1.22)$ & $(3.0,\ 8.7,\ 8.7)$\tabularnewline
\hline
$(+,-)$ & $(1.91,-1.89)$ & $8.2$ & $(1.46,-1.44)$ & $(0.13,-0.14)$ & $(4.0,\ 4.0,\ 4.3)$\tabularnewline
\hline
$(-,-)$ & $(0.74,-0.71)$ & $3.1$ & $(-1.49,1.48)$ & $(-1.19,1.17)$ & $(3.5,\ 7.9,\ 7.9)$\tabularnewline
\hline
$(-,-)^*$ & $(0.74,-0.71)$ & $3.1$ & $(-1.49,1.48)$ & $(-1.19,1.17)$ & $(3.5,\ 7.8,\ 8.0)$\tabularnewline
\hline
\end{tabular}
\caption{The eight possible solutions to the RH neutrino compact
  spectrum conditions of Case~1: $A_{23}=A_{33}=0$. The real
  parameters $m_{1}$ and $M_i$ also come in pairs as the complex
  phases $\delta,\,\alpha$ and $\beta$. A single approximate value is
  displayed because the difference between the two values is
  numerically irrelevant.  For the $(+,-)$ solution, the almost
  degenerate RH neutrinos are the lighter ones, while in all the other
  cases are the heavier ones.
  In the last line, labeled
  with an asterisk $(-,-)^*$, we give the results for the $(-,-)$
  solution in which   $A_{23}$ and $A_{33}$ are set to  arbitrary
  small, but nonvanishing values (see text).
}
\label{tab:2}
\end{center}
\end{table}
The conditions $A_{23}=A_{22}=0$ of Case 2 have only 2+2 solutions,
that are listed in Table~\ref{tab:3}.  In this case there is a very
large hierarchy ${\cal O}(10^7)$ between the two almost degenerate RH
neutrino masses $M_{1,2}$ and $M_3$ so that, as expected, the RH
spectrum is not compact.  Perturbativity of the Yukawa couplings then
implies $\Lambda_R\gsim 10^{14}\,$GeV. Accommodating such a large
intermediate scale might be problematic in the SO(10) model.
Furthermore, as we will see in the next section, in this case
leptogenesis is unsuccessful for both the $(-,+)$ and $(-,-)$
solutions.

To conclude this section, we have seen that by forcing the SO(10)
model to produce a compact RH neutrino spectrum we obtain a scenario
in which all the parameters relevant for leptogenesis remain
determined in terms of the set of low energy observables that have
been already measured. Most remarkably, as we will see, besides
predicting values for the absolute neutrino mass scale $m_1$, the CP
violating phases, and the RH neutrino spectrum, the size {\it and the
  signs} of the flavoured CP asymmetries relevant for leptogenesis are
also fixed, and allow to predict the size {\it and sign} of the
cosmological baryon asymmetry generated through leptogenesis. Such a
level of predictability is indeed quite unusual in see-saw inspired
scenarios. Clearly, verifying if the baryon asymmetry yield of
leptogenesis is in agreement with observations will now represent the
major test of our scenario. This is the task that we are going to
address in the next section.

\begin{table}[h!]
\begin{center}
\renewcommand{\arraystretch}{1.3}
\begin{tabular}{|@{\;}c@{\;}|@{\;}c@{\;}|@{\;}c@{\;}
|@{\;}c@{\;}|@{\;}c@{\;}|@{\;}c@{\;}|@{\;}c@{\;}|}
\hline
$\left(\theta_{12},\theta_{23}\right)$&$\delta$
&$m_{1}\,$\small($10^{-3}\,$eV)
&$\alpha$&$\beta$
&$M_{1,2}\,$\small($10^{7}\,$GeV)
&$M_{3}\,$\small($10^{14}\,$GeV)
\tabularnewline
\hline
\hline
$(-,+)$ & $(2.66,-2.69)$ & $2.0$ &
$(-1.51,1.52)$ & $(-0.21,0.23)$ & $3.1$ & $3.6$\tabularnewline
\hline
$(-,-)$ & $(0.35,-0.32)$ & $2.0$ & $(1.53,-1.54)$ & $(0.19,-0.21)$
& $3.3$ & $3.0$
\tabularnewline
\hline
\end{tabular}
\caption{Same than Table~\ref{tab:2},  for the  four possible
solutions to the conditions of Case~2:
$A_{23}=A_{22}=0$.}
\label{tab:3}
\end{center}
\end{table}
\bigskip

\section{Leptogenesis}
\label{LEPTO}

In the basis for the Dirac mass matrix as in \eqn{eq:mdtilde},
the CP asymmetry in the decay of the RH neutrino $N_i$ ($i=1,2,3$)
to a lepton $\ell_\alpha$  ($\alpha=e,\mu,\tau$) is given by~\cite{Covi:1996wh}
\begin{eqnarray}
\epsilon_{i\alpha} & = & \frac{1}{8\pi v^{2}}\sum_{k\neq i}
\frac{{\rm Im}\left[\left(\hat{m}_{D}^{\dagger}\right)_{i\alpha}
\left(\hat{m}_{D}\right)_{\alpha k}
\left(\hat{m}_{D}^{\dagger}\hat{m}_{D}\right)_{ik}\right]}
{\left(\hat{m}_{D}^{\dagger}\hat{m}_{D}\right)_{ii}}
f\left(\frac{M_{k}^{2}}{M_{i}^{2}}\right)
\nonumber \\
& + & \frac{1}{8\pi v^{2}}\sum_{k\neq i}
\frac{{\rm Im}\left[\left(\hat{m}_{D}^{\dagger}\right)_{i\alpha}
\left(\hat{m}_{D}\right)_{\alpha k}\left(\hat{m}_{D}^{\dagger}
\hat{m}_{D}\right)_{ki}\right]}
{\left(\hat{m}_{D}^{\dagger}\hat{m}_{D}\right)_{ii}}
g\left(\frac{M_{k}^{2}}{M_{i}^{2}}\right),
\label{eq:CP_asym}
\end{eqnarray}
where $v=174$ GeV is the EW vev and\footnote{For the resonant terms,
  we used the expressions from
  Refs.~\cite{Buchmuller:1997yu,Anisimov:2005hr}.}
\begin{eqnarray}
f(x) & = & \sqrt{x}\left[\frac{1-x}{\left(1-x\right)^{2}+
\left(\frac{\Gamma_{i}}{M_{i}}-x\frac{\Gamma_{k}}{M_{k}}\right)^{2}}+1-
\left(1+x\right)\log\frac{1+x}{x}\right],\nonumber \\
g(x) & = & \frac{1-x}{\left(1-x\right)^{2}+
\left(\frac{\Gamma_{i}}{M_{i}}-x\frac{\Gamma_{k}}{M_{k}}\right)^{2}},
\label{eq:f_and_g}
\end{eqnarray}
are loop functions with $\Gamma_{i} \equiv \frac{M_{i}}{8\pi v^{2}}
(\hat{m}_{D}^{\dagger}\hat{m}_{D})_{ii}$ the total $N_i$ width.  The
expression in the second line of eq.~\eqref{eq:CP_asym} corresponds to
the lepton-flavour-violating but lepton-number-conserving self-energy
diagram. It vanishes when summed over $\alpha$ so it does not
contribute in the one-flavour approximation, but plays an important
role~\cite{Nardi:2006fx} when, as in our case, leptogenesis occurs in
the flavoured regime.  In eqs.~\eqref{eq:f_and_g}, $g(x)$ and the
first term in the square bracket of $f(x)$ come from the
self-energy contributions with the resonant condition given by
\begin{eqnarray}
1-x & = & \pm\left(\frac{\Gamma_{i}}{M_{i}}-x\frac{\Gamma_{k}}{M_{k}}\right),
\label{eq:res_con}
\end{eqnarray}
while the remaining terms in $f(x)$ correspond to contributions from
the vertex diagram. The resonant condition eq.~\eqref{eq:res_con}
gives
\begin{eqnarray}
f(x) & \simeq & \sqrt{x}g(x)
=\frac{\sqrt{x}}{2\left(\frac{\Gamma_{i}}{M_{i}}-x\frac{\Gamma_{k}}{M_{k}}\right)},
\end{eqnarray}
where in $f(x)$ we have ignored the subleading contributions of the
vertex diagram.  However, in our case although the degeneracy
conditions $M_2\sim M_3$ or $M_1\sim M_2$ are approximately fulfilled,
we never reach a fully resonant regime as defined by
eq.~\eqref{eq:res_con}, and hence ignoring the ``regulator'' term
$\left(\frac{\Gamma_i}{M_i}-x\frac{\Gamma_k}{M_k}\right)$ in
eqs.~\eqref{eq:f_and_g} only yields negligible numerical
differences.

In order to calculate the baryon asymmetry, we need to solve a set of
Boltzmann equations (BE) (we refer to~\cite{leptoreview} and
references therein for details).  By including for simplicity only
decays and inverse decays, the BE for the RH neutrino densities
$Y_{N_i}$ and for $Y_{\Delta_\alpha}$, that is the asymmetry density
of the charge $B/3-L_\alpha$ normalized to the entropy density $s$,
can be written as:
\begin{eqnarray}
sHz\frac{dY_{N_{i}}}{dz} & = & -\gamma_{N_{i}}
\left(\frac{Y_{N_{i}}}{Y_{N}^{eq}}-1\right),\nonumber \\
sHz\frac{dY_{\Delta_{\alpha}}}{dz} & = &
-\sum_{i}\left[\epsilon_{i\alpha}\gamma_{N_{i}}
\left(\frac{Y_{N_{i}}}{Y_{N}^{eq}}-1\right)
-\frac{\gamma_{N_{i\alpha}}}{2}
\left(\frac{Y_{\Delta\ell_{\alpha}}}{Y_{\ell}^{eq}}+
\frac{Y_{\Delta H}}{Y_{H}^{eq}}\right)\right],
\label{eq:BE}
\end{eqnarray}
where $Y_{N}^{eq}=\frac{45}{4\pi^{4}g_{*}}z^{2}{\cal K}_{2}(z)$ is the
equilibrium density for the RH neutrinos with $g_*=106.75$ and ${\cal
  K}_{2}$ the second order modified Bessel function of the second
kind, $2Y_{\ell}^{eq}=Y_{H}^{eq}=\frac{15}{4\pi^{2}g_{*}}$ are
respectively the equilibrium densities for lepton doublets and for the
Higgs, and the integration variable is $z=M/T$ with $T$ the
temperature of the thermal bath, and $M=M_{1,2,3}$ the mass of the decaying neutrino.  In
the above, we have defined $Y_{\Delta_{\alpha}}=Y_{\Delta
  B}/3-Y_{\Delta L_{\alpha}}$ with $Y_{\Delta L_{\alpha}}$ the total
lepton density asymmetry in the $\alpha$ flavour which also includes
the asymmetries in the RH lepton singlets.  Since RH neutrinos only
interact with lepton doublets, the RHS of the second equation of
eqs. \eqref{eq:BE} involves only the LH lepton doublets density
asymmetry in a given flavour $\alpha$,
$Y_{\Delta\ell_{\alpha}}=A_{\alpha\beta}Y_{\Delta_{\alpha}}$ with
$A_{\alpha\beta}$ the flavour mixing matrix~\cite{flavour0} given in
\eqn{eq:Amatrix} below. In the same equation, we also define
$Y_{\Delta H}=C_{\beta}Y_{\Delta_{\beta}}$ the Higgs density asymmetry
with $C_{\beta}$~\cite{spectator2} also given in \eqn{eq:Amatrix} and
$\gamma_{N_{i\alpha}}=P_{i\alpha}\gamma_{N_{i}}$ (no sum over $i$)
where $P_{i\alpha}$ projects the decay rate over the $\alpha$ flavour,
that is, it corresponds to the branching ratio for $N_{i}$ decaying to
$\ell_{\alpha}$, and can be written as
\begin{equation}
P_{i\alpha}=\frac{\left(\hat{m}_{D}^{\dagger}\right)_{i\alpha}
\left(\hat{m}_{D}\right)_{\alpha i}}
{\left(\hat{m}_{D}^{\dagger}\hat{m}_{D}\right)_{ii}}.
\label{eq:fla_proj}
\end{equation}
Let us also introduce the rescaled decay width
\beq
{\widetilde m}_i \equiv
\frac{8\pi{ v^2}}{M_i^2}\Gamma{_i} =
\frac{(\hat{m}_D^\dagger\hat{m}_D)_{ii}}{  M_i},
\eeq
which is also known as the effective washout parameter,  that
parametrizes conveniently the departure from thermal equilibrium of
$N_i$-related processes (the larger ${\widetilde m}_i$, the closer to
thermal equilibrium the decays and inverse decays of $N_i$ occur, thus
suppressing the final lepton asymmetry).  Finally, the combination
$P_{i\alpha}\, {\widetilde m}_i$ projects the washout parameter over a
particular flavour direction, and determines how strongly the lepton
asymmetry of flavour $\alpha$ is washed out.

Leptogenesis becomes possible when the thermal bath temperature
approaches the value of the mass of the decaying RH neutrino, that
becomes non relativistic and can decay. However, when the washout
parameter is large $\widetilde m \gsim 10^{-3}$, at $T\sim M_i$ the RH
neutrinos are actually in equilibrium and no asymmetry can be
generated.  In our case the RH neutrinos are coupled rather strongly
to the thermal bath ($\widetilde m \gg 10^{-3}$ see
e.g. Table~\ref{tab:4}) and in this case the generation of the bulk of
the lepton asymmetry is delayed down to much lower temperatures: for
$\widetilde m \sim 6\times 10^{-2}$ for example one can estimate
$z=M/T\sim 8$~\cite{Buchmuller:2004nz}).
Thus, the range of temperatures where the lepton asymmetry is
generated falls well below $10^{9}\,{\rm GeV}$, where both
the $\tau$ and $\mu$ Yukawa interactions are presumably in
equilibrium~\cite{leptoreview}. In this regime all the three lepton
flavours are then distinguished, and their dynamical evolution must
be followed separately.
The $A$ flavour mixing matrix and the $C$ vectors allow
to accomplish this task, and in our temperature regime are given
by~\cite{Nardi:2006fx}
\begin{eqnarray}
A & = & \frac{1}{2148}\left(\begin{array}{ccc}
-906 & 120 & 120\\
75 & -688 & 28\\
75 & 28 & -688\end{array}\right),\nonumber \\
C & = & -\frac{1}{358}\left(37,52,52\right).
\label{eq:Amatrix}
\end{eqnarray}
Once the final asymmetries in the lepton flavour charge densities
$Y_{\Delta_{\alpha}}$ are obtained by solving numerically the
BE~\eqn{eq:BE}, the baryon asymmetry generated through leptogenesis is
given by~\cite{Harvey:1990qw}
\begin{eqnarray}
Y_{\Delta B} & = & \frac{28}{79}\sum_{\alpha}Y_{\Delta_{\alpha}}\,.
\label{eq:Y_B}
\end{eqnarray}
The resulting prediction should then be confronted with the
experimental number. The most precise experimental determination of
$Y_{\Delta B}$ is obtained from measurements of the cosmic microwave
background (CMB) anisotropies.  A fit to the most recent observations
(WMAP7 data only, assuming a $\Lambda$CDM model with a scale-free
power spectrum for the primordial density
fluctuations)~\cite{Larson:2010gs}, when translated in terms of
$Y_{\Delta B}$ gives at 95\% c.l.~\cite{Fong:2011yx}
\begin{equation}
\label{eq:YB_CMB}
Y_{\Delta B}^{CMB}=(8.79 \pm 0.44) \times 10^{-11}.
\end{equation}

\subsection{Numerical Results}

For all the phenomenologically viable cases stemming out from our
scenario, the complete set of high energy parameters required for
computing the baryon asymmetry yield of leptogenesis is predicted.
The RH neutrino masses are listed for Case 1 in the last  column
in Table~\ref{tab:2}, and for Case 2 in the last  column of
Table~\ref{tab:3} (of course, the precise numerical values of the
masses, rather then the approximate values listed in the tables, are
used for the leptogenesis computation). As regards the flavoured CP
asymmetries, they are computed according to~\eqn{eq:CP_asym} and the
corresponding results for our two cases are listed in
Table~\ref{tab:4} and~\ref{tab:5}.  Case 1 (Table~\ref{tab:4}) results
in a very compact spectrum of RH neutrinos
(and  with a pair of almost degenerate states
if the exact conditions $A_{23}=A_{33}=0$ are imposed).
For solutions $(+,+)$ and $(+,-)$ the largest mass differences remain
at the 10\% level, while for solutions $(-,+)$ and $(-,-)$ they reach
a factor of a few. Under this conditions it is mandatory to include
the contributions from the heavier RH neutrinos, that in the first
case (of tiny mass differences) can affect the results for a factor up
to ${\cal O}(10^3)$.  For Case 1 (Table~\ref{tab:4}), that contains
sub-cases in which leptogenesis can be successful, we give the
complete list of the $N_1$ flavoured parameters: the CP asymmetries
for the three flavours are given in columns 2-4, the washout flavour
projectors in columns 5-7, and the effective washout parameters in
column 8. The final results obtained by integrating the BE and by
converting $Y_{\Delta_{B-L}}=\sum_\alpha Y_{\Delta_{\alpha}}$ into
$Y_{\Delta_{B}}$ through \eqn{eq:Y_B} are given in the last column of
the two tables.  Columns 2-4 in Table~\ref{tab:4} show that
asymmetries of different signs are produced in different lepton
flavours (an example of this situation is also depicted in the left
panel in Fig.~\ref{fig:2}), while columns 5-7 show that a certain
hierarchy exists between the flavoured washout parameters.  Given the
highly non uniform pattern in flavour space of the relevant
leptogenesis quantities it is clear that no analytical expression
based on the single flavour approximation would produce a reliable
result, and we can firmly conclude that lepton flavour dynamics is of
crucial importance for studying leptogenesis in the SO(10)
model\footnote{For example, an analysis in some aspects similar to
  ours, but in which flavour dynamics is neglected, has been carried
  out in Ref.~\cite{Akhmedov:2003dg}.  Their `Special case III' is
  similar to our Case 1, but the conclusions are opposite. This is
  most likely due to the enhancements of the leptogenesis efficiency
  from flavour effects.}.

%
\begin{table}[t!]
\renewcommand{\arraystretch}{1.3}
\begin{center}
\begin{tabular}{|@{\;}c@{\;}|@{\;}c@{\;}|@{\;}c@{\;}
|@{\;}c@{\;}|@{\;}c@{\;}|@{\;}c@{\;}|@{\;}c@{\;}
|@{\;}c@{\;}|@{\;}c@{\;}|}
  \hline
$\left(\theta_{12},\theta_{23}\right)$ & $\epsilon_{1e}\,$\small$(10^{-5})$
&$\epsilon_{1\mu}\,$\small$(10^{-5})$&$\epsilon_{1\tau}\,$\small$(10^{-5})$
& $P_{1e}$ & $P_{1\mu}$ & $P_{1\tau}$ & $\widetilde{m}_{1}$\small($10^{-2}$ eV) & $Y_{\Delta B}$\small$(10^{-10})$\tabularnewline
  \hline
  \hline
  $(+,+)$ & $(0.9,0.5)$ & $(3.4,-5.0)$ & $(-4.3,4.5)$ & $10^{-3}$ & $0.03$ & $0.97$ & $74$ & $(-0.3,-0.08)$\tabularnewline
  \hline
  $(-,+)$ & $(0.041,0.083)$ & $(-0.52,0.41)$ & $(0.47,-0.49)$ & $0.03$ & $0.50$ & $0.47$ & $6.2$ & $(-0.8,-1.5)$\tabularnewline
  \hline
  $(+,-)$ & $(-0.001,-0.001)$ & $(0.003,-0.001)$ & $(-2.5,2.6)$ & $10^{-4}$ & $10^{-3}$ & $1.0$ & $10^5$ & $(0.03,0.07)$\tabularnewline
  \hline
  $(-,-)$ & $(-0.16,-0.06)$ & $(-0.89,1.14)$ & $(1.05,-1.09)$ & $0.03$ & $0.47$ & $0.50$ & $6.7$ & $(2.3,0.32)$\tabularnewline
  \hline
  $(-,-)^*$ & $(-0.05,-0.04)$ & $(-0.05,0.15)$ & $(-0.15,0.15)$ & $10^{-3}$ & $0.06$ & $0.94$ & $70$ & $(0.43,0.38)$\tabularnewline
  \hline
\end{tabular}
\caption{The leptogenesis parameters corresponding to the different
solutions of Case 1.}
\label{tab:4}
\end{center}
\end{table}

For Case~2, given that all the solutions consistent with the low
energy constraints eventually fail the leptogenesis test, we just give
in Table~\ref{tab:5}  the total CP asymmetries and the approximate
values of the total washout parameters for the two quasi degenerate RH
neutrinos $N_{1,2}$.  This reduced set of figures is however
sufficient to conclude at a first glance that with CP asymmetries of
${\cal O}(10^{-8})$ and washout parameters of ${\cal O}(1\,$eV), no
flavour dynamics could rescue leptogenesis from a
quantitative failure.

Our results are collected in the last columns of Tables~\ref{tab:4}
and Table~\ref{tab:5}.
Although we are using somewhat simplified BE in which thermal
corrections~\cite{thermal}, scatterings and CP violation in
scatterings~\cite{flavour3,Nardi:2007jp,Fong:2010bh}, and other
subleading effects are neglected, the estimates of the final baryon
asymmetry we obtain should be sufficiently accurate for our scopes.
For example, we have checked that including scatterings and CP
violations in scatterings introduces a $\lesssim~$25 \% effect, which
is by no means crucial to test the scenario.  There are in fact
other important sources of uncertainties: in our analysis we are using
fixed central vales for all the input parameters, and it goes without
saying that the final value of $Y_{\Delta B}$ will be affected by the
experimental uncertainties.  Even more importantly, there are also
theoretical uncertainties stemming from deviations from the exact
quark-lepton symmetry ansatz~\eqn{eq:terza}, as well as from
deviations from the exact zeroes in the conditions $A_{23}=A_{33}=0$,
which are obviously difficult to quantify.
Therefore, we will be contented to require that
a successful prediction of the BAU, besides having the correct sign,
should approach the experimental result~\eqn{eq:YB_CMB} only within a
factor of a few.
For Case 1, we obtain four solutions with the wrong (negative) sign of
the BAU, and other four with the correct sign. They are listed in
Table~\ref{tab:4}. However, only the two solutions in the fourth line
of the Table are sufficiently close to the experimental
value~\eqn{eq:YB_CMB} to be all phenomenologically acceptable.  For
this two solutions we give in the last line of the Table the values of
the leptogenesis parameters for $N_1$, and of the final baryon
asymmetry when the exact zeroes in \eqn{eq:conditions} are lifted to
small but not vanishing values as given in \eqn{eq:nonzero}.  We see
that although the final value of $Y_{\Delta B}$ is sensitive to this
change, it still remains within a factor of two from the measured
central value~\eqn{eq:YB_CMB}.

As we have already said, in this SO(10) scenario the leptogenesis
efficiency gets largely enhanced by flavour effects, and it is then
worth asking what would happen if the bulk of the lepton asymmetry,
rather than in the three flavour regime, is generated when only the
$\tau$ Yukawa coupling mediated in-equilibrium reactions, and the
number of relevant flavour is reduced to two. In the two flavour and
strong washout case, estimating $Y_{\Delta B}$ is more subtle because
there can be protected directions in which the asymmetry generated by
$N_{2,3}$ is not erased by $N_1$ washouts~\cite{N2,Antusch:2010ms}.
We follow a simplified approach that neglects this phenomena, and thus
gives a conservative estimate of the final asymmetry, obtaining for the $(-,-)$
case $Y_{\Delta B}\gsim (0.3,-0.3) \times 10^{-10}$.  In the first
case the asymmetry gets reduced, but still remains within a factor of
three from the experimental number; however, in the second case the
asymmetry changes sign, which again shows the importance of a proper
treatment of flavour dynamics.

As regards Case 2, we see that the two right-sign solutions yield a
baryon asymmetry that is too small by almost three order of
magnitudes. This suppression of $Y_{\Delta B}$ is due
to two different reasons: firstly the CP asymmetries are
exceedingly small because the  imaginary parts of the relevant combinations
of couplings are strongly suppressed, and secondly the washout parameters
are rather large, and imply an almost in-equilibrium dynamics that impedes
building up any sizable density asymmetries.

In conclusion, the SO(10) model constrained by the assumption of the
quark-lepton symmetry in~\eqn{eq:terza} and by the compact RH neutrino
spectrum conditions in~\eqn{eq:conditions}, when confronted with the
results from neutrino oscillation experiments, and with the
requirement of successful leptogenesis, yields predictions for the yet
unknown low energy neutrino parameters, that are summarized in the
following two possibilities
\begin{eqnarray}
  \label{eq:results}
 &&
(--):  \qquad
m_{\nu_1} \simeq 3\times 10^{-3}\,{\rm eV}, \qquad
  \delta\simeq  \,\pm\, 0.7\,, \qquad
\alpha \simeq \,\mp 1.5\,, \qquad
\beta\simeq \,\mp 1.2\,,
\end{eqnarray}
which correspond to either the upper or lower sign of the three
phases.

With the numerical results listed in~\eqn{eq:results} another low
energy observable can be predicted, that is the neutrinoless
double beta decay 
effective parameter
\begin{equation}
m_{ee} \equiv (m_\nu)_{11} = \sum_i
({U_{PMNS}^\star})_{1i} {(m_\nu)_i^{diag}} ({U_{PMNS}^\dagger})_{i1}\,,
\label{eq:0n2b}
\end{equation}
for which we obtain $|m_{ee}|\lsim 2 \>\times 10^{-3}\,$eV that, as
could have been expected for hierarchical and normal ordered neutrino
masses, remains well below the sensitivity of all ongoing and planned
experiments~\cite{Barabash:2011fg}.

\begin{table}[t!]
\renewcommand{\arraystretch}{1.3}
\begin{center}
\begin{tabular}{|c|c|c|c|c|}
\hline
$\left(\theta_{12},\theta_{23}\right)$&$\epsilon_{1}\,$\small$(10^{-8})$
& $\epsilon_{2}\,$\small$(10^{-8})$ & $\widetilde{m}_{1,2}\,$\small(eV)
& $Y_{\Delta B}\,$\small$(10^{-14})$\tabularnewline
\hline
\hline
$(-,+)$ & $(5.4,-5.5)$ & $(5.4,-5.5)$ & $1.7$ & $(-1.8,1.8)$\tabularnewline
\hline
$(-,-)$ & $(-4.3,5.0)$ & $(-4.3,5.0)$ & $1.6$ & $(1.5,-1.7)$\tabularnewline
\hline
\end{tabular}
\caption{The leptogenesis parameters corresponding to the different
  solutions of Case 2.}
\label{tab:5}
\end{center}
\end{table}

\section{Discussion and Conclusions}
\label{CONCLU}

The predictive power of our scenario is spelled out in clear in
\eqn{eq:results}, and such a high level of predictability calls for an
explanation.  The crucial point is that in our study there are no free
parameters: everything is fixed in terms of the low energy neutrino
observables and by the additional assumption of quark-lepton
symmetry~\eqn{eq:terza} and by the compact RH spectrum conditions
$A_{23}=A_{33}=0$.  The only freedom left over by these latter
constraints is a discrete one, and corresponds to the signs of the two
angles $\theta_{12}$ and $\theta_{23}$, for each choice of which there
are in turn two solutions, corresponding to positive and negative
values of the phase $\delta$.
Given that there is no free parameter
that can be adapted to fit the observed value of the BAU, we find
intriguing that among the discrete set of eight possibilities of Case
1,
in two cases the leptogenesis yield of baryon asymmetry is
in acceptable agreement with observations.

To summarize  the main results of the paper, we have first shown that in
the SO(10) seesaw model it is technically possible to arrange for a
compact RH neutrino spectrum, and this in spite of the fact that the
SO(10) neutrino Dirac mass matrix is characterized by a hierarchy
between its eigenvalues that is much stronger than the one observed
for the the light neutrinos, a situation that would naturally call for
a compensating large hierarchy in the RH masses.  We have argued that
this possibility can be implemented in a consistent way only if the
PMNS mixing angle $\theta_{13}$ is nonvanishing, since only in this
case we have at disposal the Dirac phase $\delta$ as an additional
free physical parameter that can cope with satisfying the compact RH
spectrum conditions.  The counting of free parameters is a subtle
point: clearly our construction relies quantitatively on the
assumption of a strict quark-lepton symmetry $m_D=m_u$ and
$V_L=V_{CKM}$, and one might argue that even in the absence of
$\delta$ one could be able to find solutions  by modifying these
assumptions. Nevertheless, in SO(10) $m_D$ and $V_L$ are intrinsically
related to $m_u$ and $V_{CKM}$, simply by the fact that all fermions
of each family, including the RH neutrino, are assigned to the same
irreducible representation of the group.  The important point is that
while the precise form of the quark-lepton duality relations can be
changed according to the amount of (family dependent) contamination in the EW breaking
sector from $\mathbf{10}$ and $\mathbf{126}$ vevs, still $m_D$ and
$V_L$ cannot be regarded as independent from the corresponding
quantities in quark sector, since for any fixed pattern of vevs, some
specific relation between them remains fixed.
Therefore, the compact spectrum conditions together with any specific
assumption about lepton-quark Yukawa relations, always yields a scenario
where  the values of the yet unknown neutrino parameters can	be
predicted directly in terms of known  quantities.
This is not so for the absolute neutrino mass scale $m_{1}$ which is a true free
parameter, since it is essentially determined by the ratio
$\Lambda_{EW}/\Lambda_{R}$ where the scale $\Lambda_R$ is free, nor
for the PMNS phase $\delta$ since the complex phases in $M_R$, that
are unrelated to $V_{CKM}$ and thus are also free, concur to determine
its value.

As regards the compact RH spectrum conditions $A_{23}=A_{33}=0$
\eqn{eq:conditions}, they should be understood with a grain of salt.
Rather than corresponding to exact zeroes, the assumption is that to a
good approximation the values of these entries are negligible.
This is spelled out in \eqn{eq:compact} and \eqn{eq:hierarchy}. We
have also tested the effects of lifting the exact zeros to the small
nonvanishing values given in \eqn{eq:nonzero} finding that (i) the
quasi degeneracy in the RH neutrino mass eigenvalues is removed,
resulting in a generic compact spectrum; (ii) the predictions for the
measurable low energy parameters (the absolute neutrino mass scale and
the CP violating phases) are not changed; (iii) the effects on the
final value of $Y_{\Delta B}$ remain under control.
We stress again that we have not put forth any theoretical explanation,
as for example a symmetry argument, for why the entries $A_{23}$ and
$A_{33}$ should be particularly suppressed, and this implies that the
fact that the four (real plus imaginary) conditions can be fulfilled
only if specific quantitative relations between
$m_{1},\,\delta,\,\alpha$ and $\beta$ are satisfied, should not be
confused with a parametric (i.e. functional) dependence like
$\delta=\delta(m_{1})$ or $\alpha=\alpha(m_{1},\delta)$, which is
something that our SO(10) scenario certainly does not give, but should
rather be regarded as numerical accidents.

It is likely that in the not too far future the values of $m_{1}$ and
of $\delta$ will eventually be measured, and therefore the specific
scenario we have been exploring, and whose predictions are summarized
in \eqn{eq:results}, is straightforwardly falsifiable.  Of course, by
modifying the form of the quark-lepton symmetry relations one would
obtain numerically different predictions.  However, any different
assumption would result in the same level of predictability, and in
particular it will have to pass the leptogenesis test which, as we
have seen, is a highly nontrivial requirement. It is certainly
conceivable a situation in which no assumption will be able to
reproduce the measured values of $m_{1}$ and $\delta$ while
simultaneously pass the leptogenesis test.  We can then conclude that
the leptogenesis scenario based on SO(10) with a compact RH neutrino
spectrum is a testable physical hypothesis.

\section*{Note added}

After this paper was published in the {\tt arXiv.org}
database~\cite{Buccella:2012kc}, the Daya Bay reactor neutrino
experiment announced the measurement of a non-zero value for the
neutrino mixing angle $\theta_{13}$ with a significance of 5.2
standard deviations: $\sin^2 2\theta_{13} = 0.092 \pm 0.016 {\rm
  (stat.)} \pm {\rm 0.005(syst.)}$~\cite{An:2012eh}.  In April 2012,
the RENO experiment also reported a non zero value $\sin^2
2\theta_{13} = 0.113 \pm 0.013 {\rm (stat.)} \pm {\rm 0.019
  (syst.)}$~\cite{Ahn:2012nd} consistent with the Day Bay result.
As we have explained in the paragraph
above~\eqn{theta13}, $\theta_{13}\neq 0$ is a mandatory condition for
the consistency of our scenario, which is now ensured by the Daya Bay
and RENO results.  The experimental central values
$\theta_{13}=8.8^\circ$ (Daya Bay) and $\theta_{13} = 9.8^\circ$
(RENO) are larger than our reference value $\theta_{13} =
5.6^\circ$~\cite{GonzalezGarcia:2010er} (see below~\eqn{theta13}), and
this will slightly change the numbers in~\eqn{eq:results}. However,
the conclusions of the leptogenesis analysis that are based on the sign
of the baryon asymmetry and on its value only within a factor of a few,
will not be changed. To give an  example, for $\theta_{13}=8.8^\circ$ we obtain for the first
 of the two cases labeled $(-,-)$ in Table~\ref{tab:4} $Y_{\Delta B}=
2.9\times 10^{-10}$ instead than $Y_{\Delta B}= 2.3\times
10^{-10}$  which  clearly  implies the same  conclusions.

\section*{Acknowledgments}

We thank Luis Oliver for very interesting discussions and useful
comments on the manuscript.  G.R. acknowledges that this material is
based upon work supported in part by the National Science Foundation
under Grant No. 1066293 and the hospitality of the Aspen Center for
Physics.

\end{document}